\documentclass[12pt,preprint]{aastex}

\newcommand{\tot}{_{\rm tot}}
\newcommand{\keV}{\rm keV}

\shorttitle{Beam gyrosynchrotron emission} \shortauthors{Altyntsev et al.}

\begin{document}


\title{Broadband microwave burst produced by electron beams}

%pulse on 10 March 2002 - gyrosynchrotron emission
%generated by electron flux with anisotropic velocity distribution?}

 %[today]

\author{A.T. Altyntsev\altaffilmark{1,2}, G.D. Fleishman\altaffilmark{3,4},
G.-L. Huang\altaffilmark{1}, and  V.F. Melnikov\altaffilmark{5}
 }

\altaffiltext{1}{Purple Mountain Observatory, Nanjing 210008, China}
\altaffiltext{2}{ISZF, Institute of Solar Terrestrial Physics,
Irkutsk 664033, Russia} \altaffiltext{3}{New Jersey Institute of
Technology, Newark, NJ 07102} \altaffiltext{4}{Ioffe
Physico-Technical Institute, St. Petersburg 194021, Russia}
  \altaffiltext{5}{Radiophysical Research
Institute, Nizhny Novgorod 603950, Russia}

\date{} %\today}

\begin{abstract}
Theoretical and experimental study of fast electron beams attracts a
lot of attention in the astrophysics and laboratory. In the case of
solar flares the problem of reliable beam detection and diagnostics
is of exceptional importance. This paper explores the fact that the
electron beams moving oblique to the magnetic field or along the
field with some angular scatter around the beam propagation
direction can generate microwave continuum bursts via
gyrosynchrotron mechanism. The characteristics of the microwave
bursts produced by beams differ from those in case of isotropic or
loss-cone distributions, which suggests a new tool for quantitative
diagnostics of the beams in the solar corona. To demonstrate the
potentiality of this tool, we analyze here a radio burst occurred
during an impulsive flare 1B/M6.7 on 10 March 2001 (AR 9368,
N27W42).  The burst is remarkably suitable for this goal because of
its very short duration, wide frequency band and unusual
polarization being in the ordinary wave mode in the optically thin
range of the spectrum. Based on detailed analysis of the spectral,
temporal, and spatial relationships, we obtained firm evidence that
the microwave continuum burst is produced by electron beams. For the
first time we developed and applied a new forward fitting algorithm
based on exact gyrosynchrotron formulae and employing both the total
power and polarization measurements to solve the inverse problem of
the beam diagnostics. We found that the burst is generated by a
oblique beam in a region of reasonably strong magnetic field ($\sim
200-300$ G) and the burst is observed at a quasi-transverse viewing
angle. We found that the life time of the emitting electrons in the
radio source is relatively short, $\tau_l \approx 0.5$ s, consistent
with a single reflection of the electrons from a magnetic mirror at
the foot point with the stronger magnetic field. We discuss the
implications of these findings for the electron acceleration in
flares and for beam diagnostics.

\end{abstract}


\section{Introduction}

Electron beams are believed to represent one of  important
elementary ingredients of the solar activity \citep[][and references
therein]{Aschw_2002}. They ionize and excite hydrogen atoms in the
chromosphere giving rise to optical H$_\alpha$ flares. They are
capable of producing nonthermal hard X-ray (HXR) and gamma-ray
radiation via the bremsstrahlung mechanism as well as of driving the
chromospheric evaporation. Then, they can drive various kinetic
instabilities in the corona giving rise to a variety of coherent
radio emission types widely observed throughout the entire radio
band \citep{Aschw_2005}.

Nevertheless, there is apparent lack of the observational tools now
for quantitative diagnostics of the electron beams in the solar
atmosphere. Currently, H$_\alpha$, HXR, and gamma emissions, as well
as type III bursts in the radio range are considered to represent
signatures of electron beams. However, these processes do not offer
any reliable straightforward diagnostics of the angular
distributions of electron beams. Even though a valuable information
of the electron beam properties might in principle be derived from
the linear polarization of the H$_\alpha$ and H$_\beta$ lines in the
chromospheric flares \citep{Henoux_etal_2003}, the measurement of
linear polarization in these spectral lines is extremely difficult,
so the corresponding beam diagnostic has not been established yet
\citep{Bianda_etal_2005}. In case of HXR and gamma emission, no
method has been proposed yet to get the angular distribution of fast
electrons from the data. Moreover, many HXR bursts originate from
coronal rather than chromospheric sources
\citep{Veronig_Brown_2004}, although even in case of chromospheric
sources the HXR emission can originate from a precipitating fraction
of approximately isotropic electron distributions rather than from
beams. Finally, any diagnostics based on the type III bursts is
difficult because for a coherent process the dependence of the
output radiation on the electron beam properties is highly nonlinear
and difficult to disentangle \citep{Aschw_2002}. On the other hand,
the beam can be invisible via the type III emission if it propagates
in a dense plasma, where high collisional damping rate quenches the
beam instability and no coherent emission is generated. On top of
this, the observed fast drift of radio fine structures is not
necessarily related to the beam propagation: it can be provided by
dynamics of MHD and/or reconnection processes as well
\citep{Altyntsev_etal_2007}, while lower values of the drift rates
can be ascribed to emissions originating at thermal conduction
fronts \citep{Farnik_Karlicky_2007}. We can conclude that available
tools are currently insufficient for reliable detection and detailed
diagnostics of the beams in the solar corona.

Curiously, one of the most promising methods of the beam study,
namely, analysis of the beam-produced microwave gyrosynchrotron
radiation remains entirely unexplored yet, although the synchrotron
radiation by nonthermal electrons has long ago been recognized to be
the main mechanism producing the microwave emission in solar flares
\citep{Ramaty_1969, Ramaty_Petrosian_1972, Benka_Holman1992, BBG,
Nindos_etal_2000, Kundu_etal_2001, Trottet_etal_2002, Bastian_2006,
Gary_2006, Nindos_2006}. Exact expressions for the gyrosynchrotron
emission and absorption coefficient \citep{Eidman1958,
Eidman1959,Ramaty_1969,Ramaty94} are cumbersome and difficult for
direct use. Thus, significant efforts have been made to find simple
analytical approximations for both the synchrotron emission in the
ultrarelativistic case
\citep{Ginzburg1951,KorchakTerletsky1952,Getmantsev1952,Ginzburg1953,Korchak,GinzburgSyrovatsky1964}
and the gyrosynchrotron emission generated by nonrelativistic and
moderately relativistic electrons \citep{Dulk82, Dulk_1985, Zhou99}.
Also, simplified numerically fast computation schemes were developed
\citep{Petrosian1981,Klein1987}. All these studies, however, assumed
isotropic electron distributions or only weakly anisotropic in some
cases. These approximations are evidently insufficient to describe
and analyze the beam-produced gyrosynchrotron emission, where the
pitch-angle anisotropy is expected to be strong.

Recently, \cite{Fleish03} discovered significant effect of the
pitch-angle anisotropy on the gyrosynchrotron spectrum and
polarization. So far \citep[see for a review][]{Fl_Nobe_2006},
analysis of the microwave continuum bursts provided ample evidence
for  the loss-cone particle distribution formed due to trapping of
the accelerated electrons in the coronal magnetic loops
\citep{Melnikovetal2002a,Melnikov2006,FlGaryNita2003,Fl_etal_2007}.
In contrast, here we present a different class of events, when there
is a beam-like anisotropy of the particle distribution.

In case of the beam-like angular distributions of the fast
electrons, in particular, the high-frequency spectral index depends
noticeably on the anisotropy and the angle of view in addition to
the standard dependence on the energy distribution. The degree of
polarization can differ strongly from that in the isotropic case.
Remarkably, the sense of polarization can correspond to the
\emph{ordinary} wave mode (O-mode) in the \emph{optically thin}
range of the spectrum in contrast to X-mode polarization in the
isotropic case \citep{Fleish03}.

Thus, the beam-like anisotropy when present must be properly taken
into account for correct modeling the solar microwave continuum
bursts, which is especially important to interpret the polarization
spectra. The goal of our study is to identify an example of the
microwave burst in which the presence of electron beams is likely
and then evaluate the properties of the pitch-angle distribution and
possibly other relevant parameters of the source from the forward
fitting of the gyrosynchrotron formulae to the observed radio data.

The guess criteria for the candidates to the beam-produced microwave
bursts are the presence of either short (of the order of second)
broadband pulses or type III like drifting bursts at lower
frequencies or both. Among many burst-candidates we eventually
selected the event of 10 March, 2001. This selection is based
on fortuitous combination of radio observations of this burst made
by different observatories as well as availability of other
important context observations for this event.

Below we describe the key observational characteristics of the
event, then suggest semi-quantitative interpretation of the data,
then describe a specially developed nonlinear chi-square
minimization fit and apply it simultaneously to total power and
polarization spectra. Eventually, the fitting yields the parameters
of the angular distribution of the radiating electrons, the viewing
angle of the emission, and the characteristic life time of the
electrons at the radio source.

\section{Instrumentation}

The event was observed by a number of radio instruments.

The \emph{Nobeyama Radio Polarimeters} (NoRP) \citep{Polar,
Nakajima85} measure the total power  and circularly polarized
intensities (Stokes parameters I and V) at 1, 2, 3.75, 9.4, 17, 35,
and 80 (Stokes I only) GHz with time resolution as high as 0.1~s. We
applied corrections provided by the Nobeyama team for the
polarization data at 1 and 2 GHz and the intensity data at 80 GHz
available at the Nobeyama Observatory Internet archives.

Chinese \emph{Solar broadband Radio Spectrometers}
\citep[SRS,][]{Fu04} measure total flux and polarized flux at
frequencies 5.2-7.6~GHz with 20~MHz spectral and 5~ms temporal
resolution (NAOC, Huairou station), and the total flux at
frequencies 4.5-7.5~GHz with 10~MHz spectral and 5~ms temporal
resolution (Purple Mountain Observatory, PMO).

The \emph{Nobeyama Radioheliograph} (NoRH) \citep{Helio} obtains
images of the Sun at 17~GHz (Stokes I and V) and 34~GHz (Stokes I)
with 0.1~s temporal resolution. At the time of the burst the angular
resolution of the NoRH was 17~arc~sec at 17~GHz and 10~arc~sec at
34~GHz.

Fortuitously, the data from another big imaging \emph{Siberian Solar
Radio Telescope} (SSRT) \citep{Smolkov, Grechnev03} are available
for this burst. SSRT yields 1D brightness distributions at 5.7~GHz
with 14~ms temporal resolution. During observation of this burst,
the knife-edge beam of north-south linear array (SSRT/NS) was
directed at $43.53$~degree from the central solar meridian (angular
resolution of 24 arc sec). The east-west array (SSRT/EW) knife-edge
beam was directed at $-23.61$ degree angle from the central solar
meridian with the angular width of 16.5 arc sec.

HXR observations made at \emph{Yohkoh} satellite by Hard X-ray
Telescope (HXT) at four energy bands (L-band, 14-23~keV; M1-band,
23-33~keV; M2-band, 33-53~keV; H-band, 53-93~keV) and Wide-Band
Spectrometer (WBS) \citep{Yosh92} at the high-energy band
(80-600~keV) and low-energy band (20-80~keV) are available.

Additionally, context data of the SOHO/MDI magnetic field and
SOHO/EIT are available for the flare.

\section{Observations}

Impulsive flare (10 March 2001, M6.7/1B) has occurred in active
region AR9368 (N27W42). NoRP  recorded an intense microwave burst
at all frequency channels (Figure~\ref{norp}, left panel). The
light curves throughout all  those frequencies except 1~GHz
display a very short (about 3 s duration) broadband peak at
04:03:39.6 UT. The pulse magnitudes were exceptionally strong  in
the range 9.4 -- 35 GHz ($> 1000$ sfu).

Analysis of this flare at different wavelengths, viz., $H_\alpha$,
HXR, SXR and radio waves was published in a number of papers.
\citet{Dingetal03} and \citet{Ding03} have classified this flare as
a white-light flare. From a good time correlation of $H_\alpha$ and
Ca II 8542 \AA\ brightenings with the peak of the microwave radio
flux, they concluded that the response in optical emission is due to
chromosphere heating by an electron beam.  \citet{Uddin04} and
\citet{Chandra06}  studied evolution of the flare active region and
associated this flare with small positive polarity region emerging
near the following negative sunspot. They distinguished two bright
$H_\alpha$ kernels connected by  a dense plasma loop with distance
of about $10^4$ km between the footpoints.

\subsection{Temporal characteristics}

The entire burst duration is rather short ($\sim 40$ s) at
frequencies above 3.75 GHz. The time profiles are remarkably
similar to each other and the duration of the prominent peak was
the same throughout the entire spectral range. The light curves at
9.4--80~GHz peaks at the same time, while that at 3.75~GHz is
delayed by a fraction of second and that at 2~GHz is delayed even
stronger (by half of second). Interestingly that the decay time
becomes slightly larger at low frequencies.

\begin{figure}
\epsscale{1.20} \plottwo{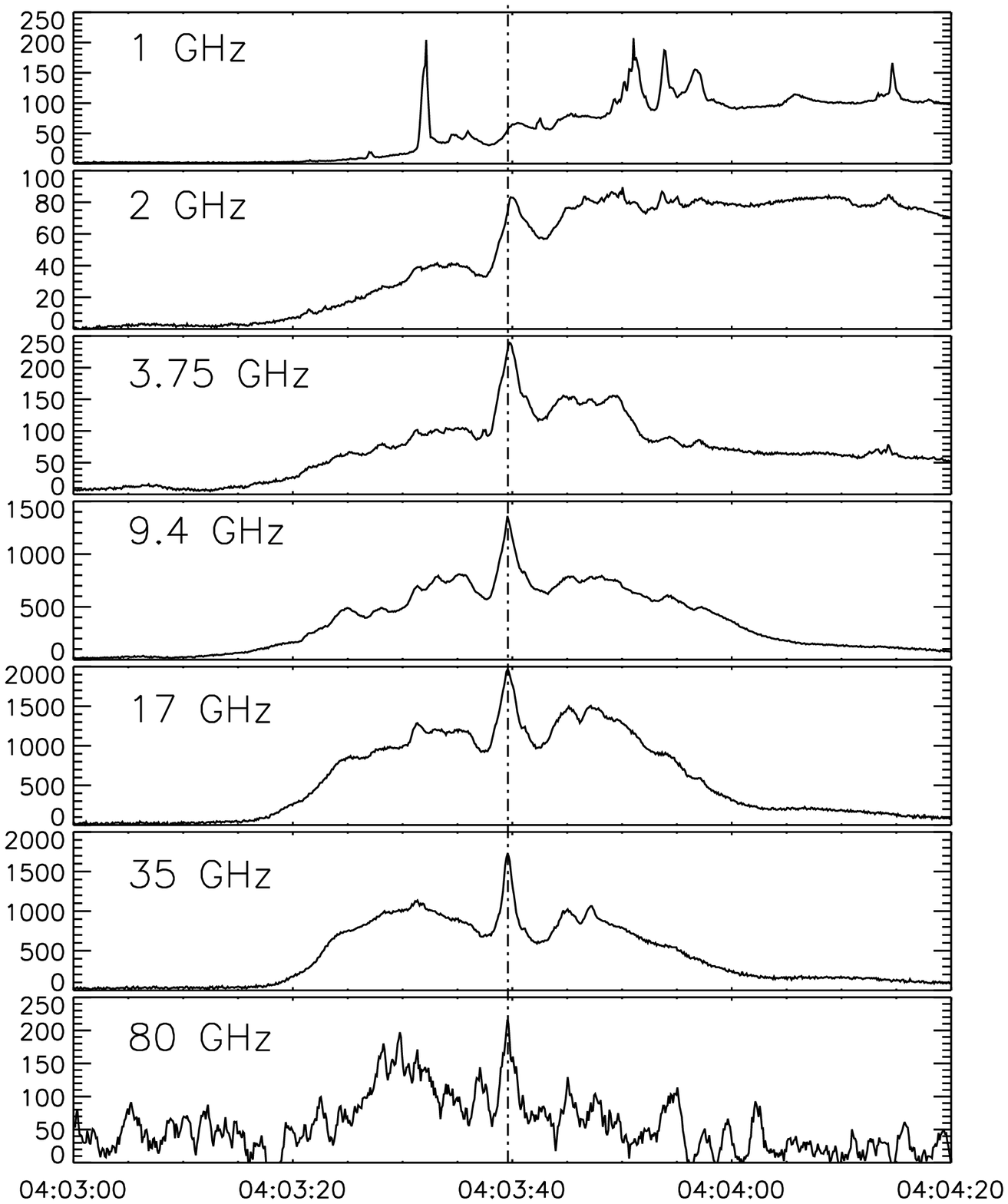}{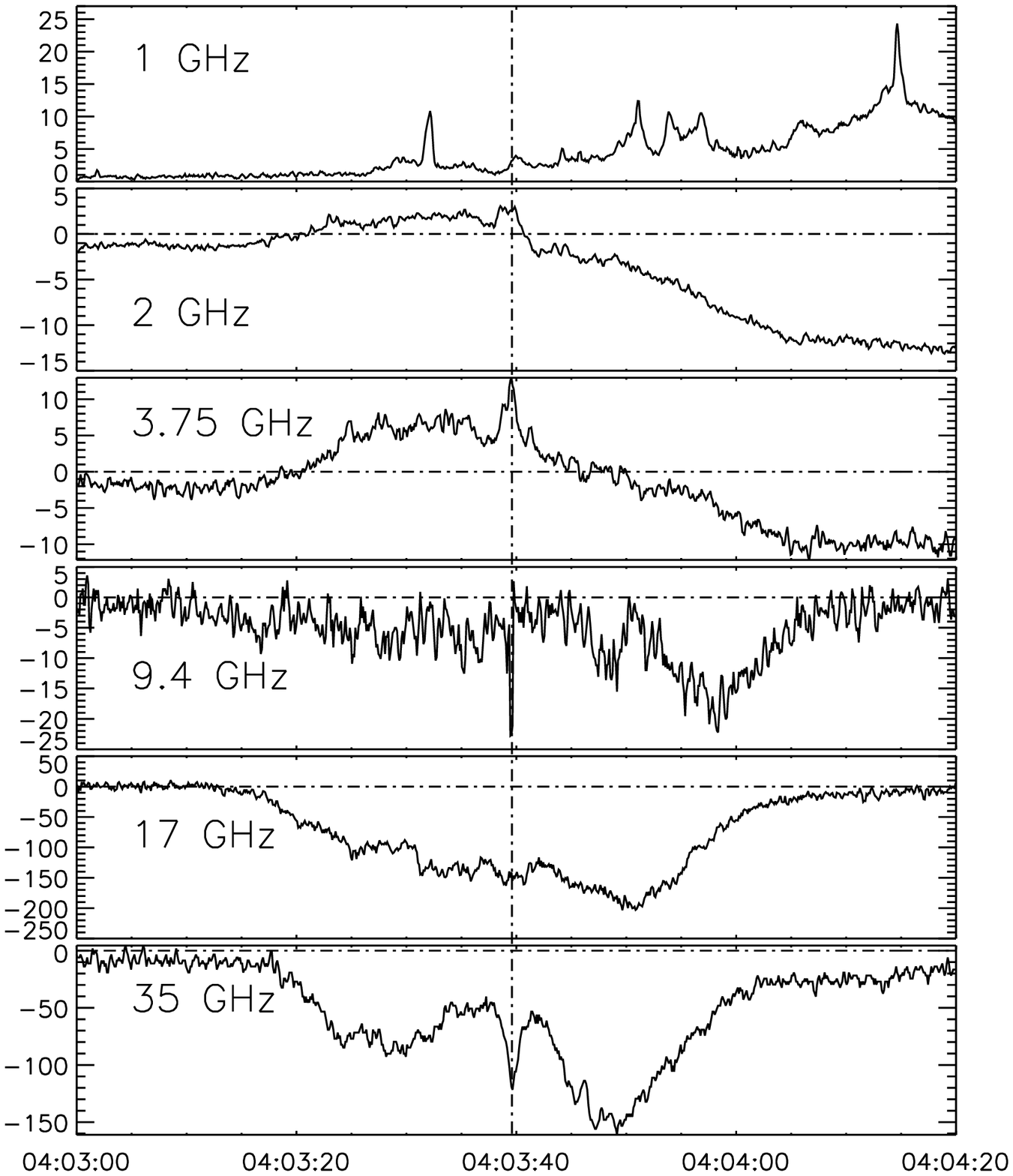} \caption{Microwave
fluxes, recorded by the NoRP polarimeters. Left: total intensity
(R+L), right:  polarized intensity (R-L). Magnitudes of fluxes are
in sfu. \label{norp}}
\end{figure}

The NoRP time profiles of the circularly polarized radiation (Stokes
V) are shown in Figure~\ref{norp} (right panel). Except 35~GHz, the
profiles of the polarized fluxes  differ essentially from the
profiles in intensity. The degree of polarization varies from
$+10\%$ at 1~GHz down to $-10\%$ at 35~GHz. The main peak is clearly
distinguished at 3.75, 9.4, and 35~GHz. Note that across the NoRP
spectrum, the right-handed circular polarization gives way to the
left-handed one somewhere between 3.75~GHz and 9.4~GHz. Analysis of
the SRS data at the range 5.2-7.6 GHz shows that this polarization
reversal occurs around 6.5 GHz.

\begin{figure}
\epsscale{1.0} \plotone{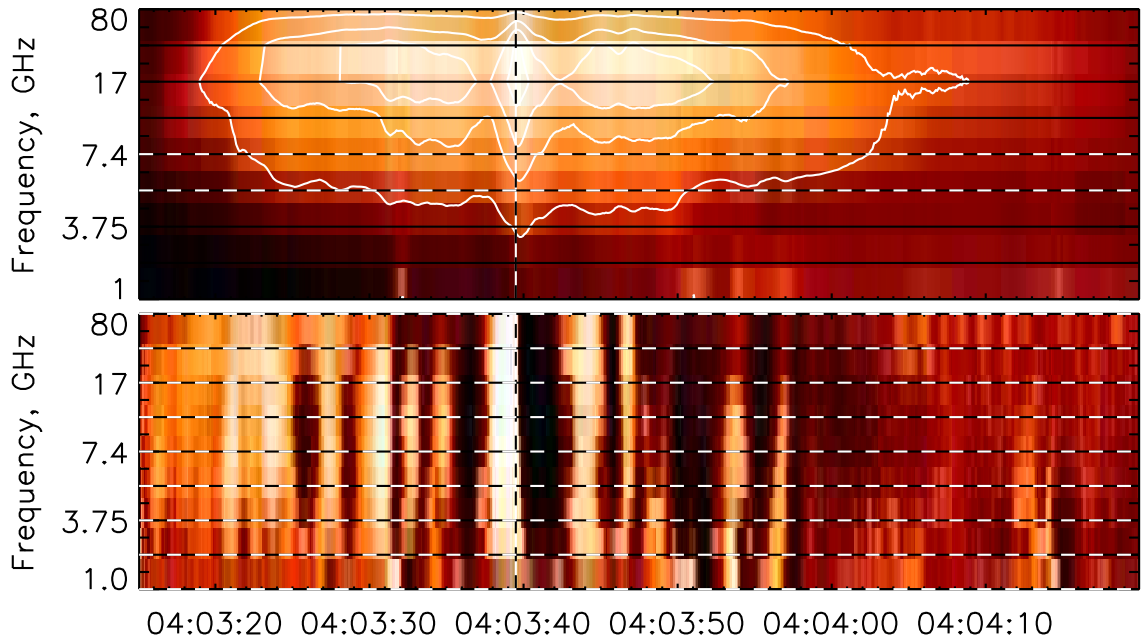} \caption{Top: Dynamic spectrum of
the 10 March 2001 flare burst. The NoRP measurements are
complemented by SRS measurements at 5.4 and 7.4~GHz. Contours: $2000
\times (0.1,0.3,0.5,0.8)$ sfu. Bottom: The derivatives of the time
profiles. \label{fig2}}
\end{figure}

In the dynamic spectrum (Figure~\ref{fig2}, top) the NoRP data are
complemented by the SRS profiles at 5.4~GHz and 7.4~GHz to fill
large gap between the NoRP receiving frequencies 3.75~GHz and
9.4~GHz.  The shape of the microwave spectrum does not change much
during the burst that is visualized by similarity of the contours of
equal intensity shown in the dynamic spectrum (Figure~\ref{fig2},
top). The time profiles in Figure~\ref{norp} show a number of pulses
besides the main peak. They are clearly seen in the time derivative
of the dynamic spectrum (Figure~\ref{fig2}, bottom).

Time derivative of dynamic spectrum (Figure~\ref{fig2}, bottom)
shows wide band fine structures during the entire  bursts. Bright
strips correspond to increase of the emission, while the dark ones
show emission decrease. The duration of the shortest strips is
comparable with the NoRP temporal resolution (0.1~s), and there
bandwidth covers the entire spectral range from 1 GHz to 80 GHz in
some cases. The strips are not noise or interference since they are
wider than several frequency channels including data from different
observatories. Note that most of the broadband pulses display no
frequency drifts, although any drift value within $\pm800$ GHz/s is
measurable with given time resolution (0.1 s) and spectral bandwidth
(80 GHz) of the pulses. The duration of decay phases (the dark
strips after the bright ones) does not depend on the frequency.

At millisecond time scale there are pulses with \emph{reverse} and
\emph{normal} frequency drift observed in the dynamic spectrum in
interval within 04:03:37 -- 04:03:56 UT, some of them are shown in
Figure~\ref{fine}. The total bandwidth of the fine structures does
not exceed 0.5 GHz and the life time is about 50 ms. The
corresponding drift rates are within $10-15$ GHz/s in this event.
The characteristic frequency of the fine structures  rises from
4.5~GHz up to 7~GHz during this interval.

\begin{figure}
\epsscale{.70} \plotone{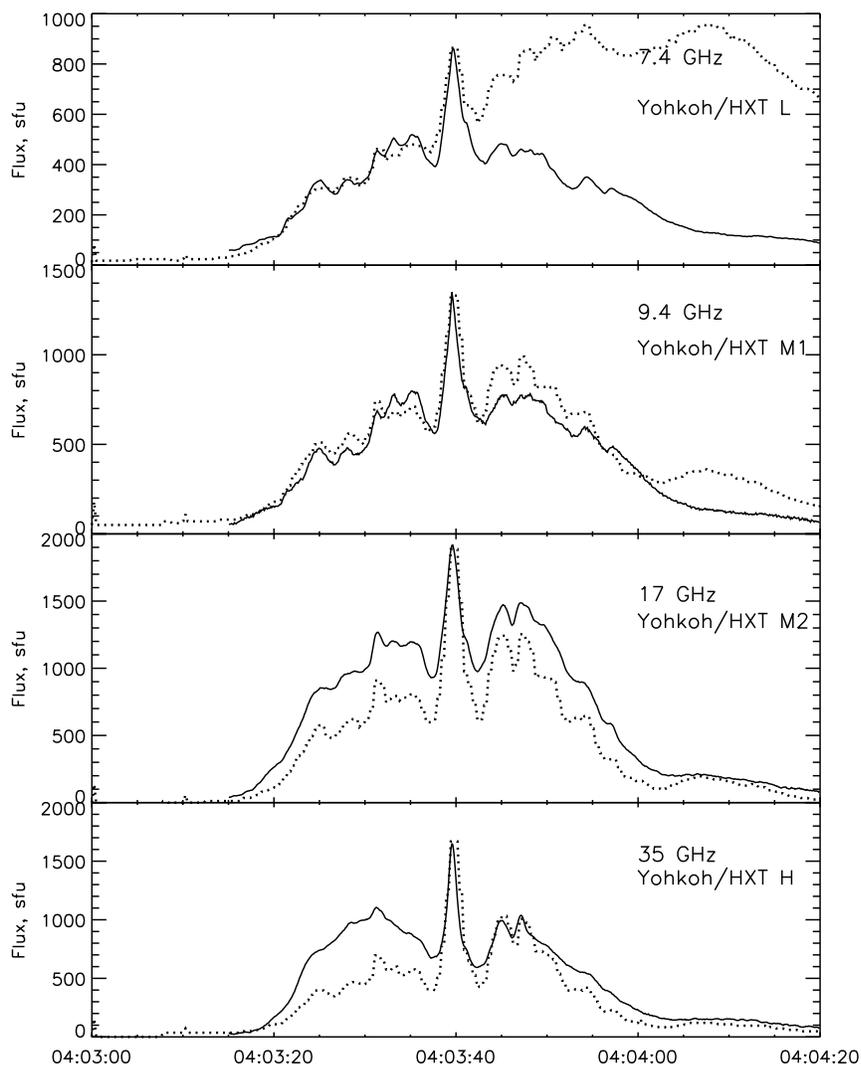} \caption{Radio vs Yohkoh/HXT light
curves. Energy ranges are L(14 –-- 23~keV), M1(23 –-- 33~keV), M2(33
--- 53~keV), H(53 –-- 93~keV). HXR light curves (dashed) are normalized such as to match
the peak value of the corresponding radio light curves (solid).
Available at
$http://gedas22.stelab.nagoya-u.ac.jp/HXT/catalogue/image\_html/eid\_html/eid\_25110.html$
\label{hxtt}}
\end{figure}

Time profiles of the hard X-ray emission are remarkably similar to
the microwave profiles (Figure~\ref{hxtt}). The same profiles were
observed with the Wide-Band Spectrometer at the high-energy band
(80-600~keV). Note, that HXR emission is delayed relative to the
radio emission. The overall L signal delays relative to various
radio frequencies by about 1-2~s, although the delay between the
impulsive peaks occurring at about 04:03:40 UT is shorter than 1 s.
At higher energies the delays are shorter being a fraction of
second. In particular, the cross correlation between the H channel
and 35~GHz light curve yields the delay of 0.5~s with the
correlation coefficient equal to $0.8$.

\subsection{Spatial  characteristics}

The flare spatial structure in $H_\alpha$ emission was studied by
\citet{Uddin04} and \citet{Chandra06}. Before the flare, emerging
flux of N polarity had penetrated into the S polarity region and
triggered the flare. From the spatial correlation between different
waveband sources it was concluded that the flare had \emph{the
three-legged} structure, i.e., it may be considered to be one of the
typical configuration of loops as determined earlier by Hanaoka
(1996). The $H_\alpha$ flare began (around 04:01 UT) as two bright
kernels (K1 and K2), which rapidly transformed during the flare peak
into a bright region expanding in the southwest direction. A short
loop connecting magnetic regions of opposite polarity is seen in 195
\AA\ emission (Figure~\ref{Fig2_3}); its length is about $10^4$ km.
One more source (so called remote source, RS) appeared after
04:03:51~UT at 17~GHz. The RS was located 150 arcsec southwest from
the main flare site (Figure~\ref{Fig2_3}) and had right handed
polarization; the presence of the RS is confirmed by a secondary
peak at 280 arc sec in the 5.7~GHz SSRT/NS scan at 04:04:21~UT
(Figure~\ref{1dssrt}, middle), which is unpolarized, however, at
this relatively low frequency.

The footpoints of the loop are seen in the 17~GHz polarization
images during the late decay phase of the burst
(Figure~\ref{pol_mag}, right panel). The shown maps are the result
of averaging of the emission over more than a minute. The averaging
is made since the brightness temperature of right handed emission
was rather weak. In the main phase of  the burst the bulk of
microwave emission is generated in the source K2 with the north
polarity of magnetic field, and it is left hand polarized. This
means that the microwave source K2 is polarized in the ordinary wave
mode. The magnetic field near the footpoints can be estimated from
the MDI magnetogram as $-170$~G (K1) and 340~G (K2).

The spatial behavior of the HXT emission is studied by
\citet{Chandra06} in detail. In all energy bands only single source
was observed. The estimated size of the HXR source was $7.0 \times
3.8$, $6.7 \times 3.3$, $6.5 \times 3.3$ and $6.2 \times 3.2$ arc
sec in L, M1, M2, and H energy bands respectively. According to
Figure ~6 in \cite{Chandra06} the HXR source is  located near source
K1.  Note that the situation when the brightness peaks in the hard
X-ray and the microwaves are located in the opposite ends of a
single loop is typical for asymmetric magnetic loops.

The SSRT 1d images are shown in Figure~\ref{1dssrt}. The vertical
lines correspond to the integration paths crossing the centers of
the two-dimensional brightness distribution at 17 GHz; these
integration paths are shown by dash-dotted lines for both EW and NS
linear arrays in Figure~\ref{Fig2_3}. The microwave source position
did not change in the NS scans, while slightly shifts to the West in
the EW scans at the burst peak. The sources K1 and K2 are not
resolved in the SSRT scans. The size of the 5.7~GHz microwave source
%scans
is about 50~arc sec in the NS scans and 25~arc sec in the EW scans.
On the contrary the polarization distributions are radically
changing with time. At the burst peak the polarization sense changes
from  the left handed to the right handed polarization and then back
in the NS scans. In the EW scans a two-polarity structure appeared
in the time interval around the burst peak.

\subsection{Spectrum description}

The microwave spectra are shown in Figure~\ref{gyro} at three
moments corresponding to the main  peak and neighboring peaks. The
spectra are similar to each other. At  low frequencies the spectra
increase with index $\gamma \simeq 1.9$. Spectral peak is at about
17~GHz. At high frequencies  the spectrum decreases rather quickly
with the high-frequency spectral index $\gamma \simeq -2.1$.

The hard X-ray spectrum $I(E)$ was studied by \citet{Chandra06}
using the Yohkoh$\backslash$HXT data. At the peak time they obtained
$I(E)=F_oE_{ph}^{-\gamma}=2.02\times 10^5 E_{ph}^{-2.4}$~photons
$\cdot$ cm$^{-2} \cdot$ s$^{-1} \cdot$ keV$^{-1}$. Assuming the
electron spectrum in the form $F_{el}(E)=AE^{-\delta}$ electrons/s
we get the parameters of electron flux under the thick-target
assumption as follows, by making necessary corrections for the
equation given by \cite{Brown}:
\begin{equation}\label{brown}
 A=5.2 \times 10^{33}\gamma^2(\gamma-1)^2B(\gamma-0.5,1.5)F_o=5.1
\times 10^{39}
\end{equation}
and $\delta=\gamma+1 =3.4$, where $E$ is expressed in keV.

\subsection{Summary of observations}

Here we summarize main observational characteristics of the event
important for further analysis:

\begin{enumerate}

\item The microwave emission in the 10 March 2001 event consists of many
short broadband pulses.

\item
Type III like features are observed around 5 GHz.

\item
The microwave emission is O-mode polarized at 17 GHz.

\item
The polarization at 5.7 GHz displays high variability and
corresponds to X-mode at the peak time.

\item
HXR light curves are remarkably similar to the microwave light
curves.

\item
HXR is delayed by a fraction of second compared with the microwave
emission.

\end{enumerate}

\section{Model}

\subsection{General trends and model dependences}

Our goal in the analysis of the microwave emission is to derive
important source parameters from forward fitting of the observed
radio spectra by the gyrosynchrotron formulae. However, as is widely
known \citep[e.g.,][]{Bastian_etal_2007} the gyrosynchrotron
emission depends on too many physical effects and parameters even if
angular distribution of fast electrons is isotropic. Expected
presence of the beam-like anisotropy of fast electrons adds a few
new free parameters, which complicates further the procedure of the
fitting. Therefore, before developing a specific forward fitting
model dealing with gyrosynchrotron emission produced by electron
beams, we critically evaluate the observed properties of the burst
to restrict or fix as many parameters as possible.

First of all, we make use of the close similarity of the radio and
HXR light curves. This similarity suggests that no trapping effect
is important for this event and both HXR and microwave emissions are
produced by the same electron distribution. The peak injection rate
of the electrons above 10 keV derived from the HXR spectrum is
$J(>10~\keV) \simeq 8.5\cdot10^{36}$ electrons/sec. We adopt that
the total number of the emitting electrons in the radio source  is
$N\tot = \tau_l J(>10~\keV)$, where $\tau_l$ is the characteristic
life time of the emitting electrons in the radio source. Since no
trapping is important, we adopt that $\tau_l$ is a single,
energy-independent, free parameter, which will be determined later
from the forward fitting model. It is clear, however, that $\tau_l$
must not exceed a few seconds, the typical duration of the single
pulses composing the burst. Then, regarding the energy dependence of
the electron distribution, we adopt the simplest assumption of a
single power-law over the momentum modulus with the spectral index
determined from the HXR spectrum (note, the energetic spectral index
of 3.4 corresponds to the index of 7.8 in the distribution over
momentum).

Second, address the question what can cause the delay of the HXRs
relative to microwave emission. If the electron beam is injected
somewhere at the top of the loop towards the foot points, then
directly precipitating electrons will first produce the HXR and a
fraction of the electrons reflected back into the loop will later
produce the radio emission. Thus, the model involving directly
precipitating beam predicts opposite delay (HXR leads radio) to what
is actually observed. Note, that models with electron trapping also
predict a delay of radio emission relative to HXR emission \citep{Melnikov1994}.

\begin{figure}
\epsscale{.70} \plotone{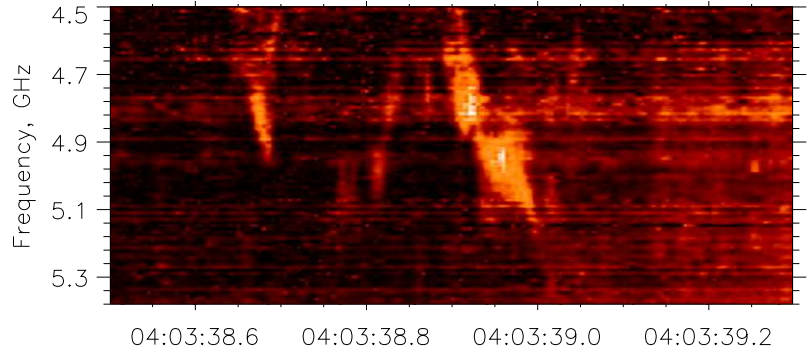}%{fine.eps}
\caption{Drifting fine structures recorded slightly before the main
flare peak with the PMO spectrometer. \label{fine}}
\end{figure}

\begin{figure}
\epsscale{1.0} \plottwo {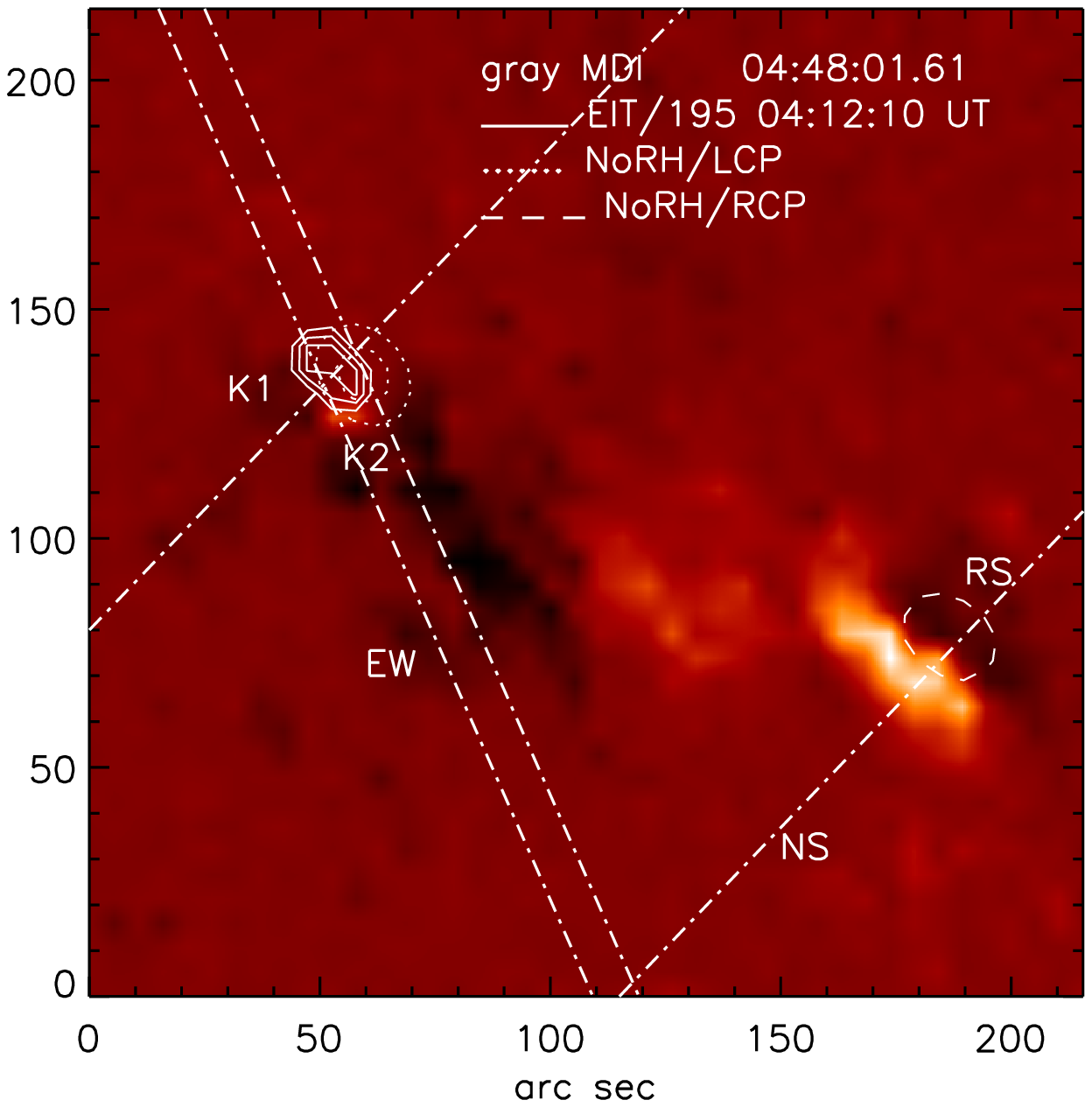}{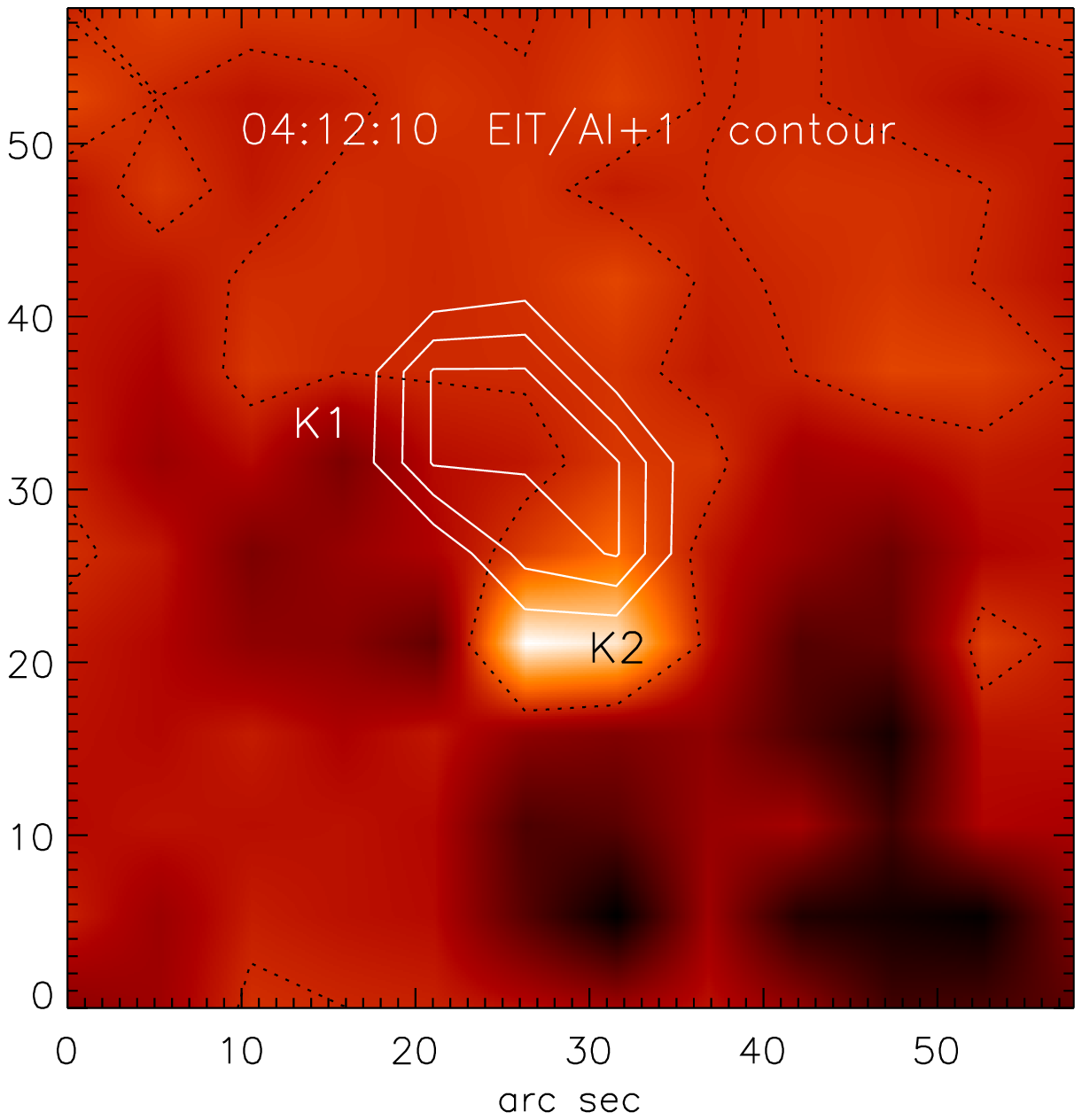} \caption{ Left: Flare
sources in UV and microwaves (right and left circular polarization).
Dash-dotted lines show directions of the EW and NS fan strips.
Right: Extended EIT partial frame with MDI magnetogram background.
Dashed contour shows the magnetic neutral line. \label{Fig2_3}}
\end{figure}

\begin{figure}
\epsscale{0.7} \plotone{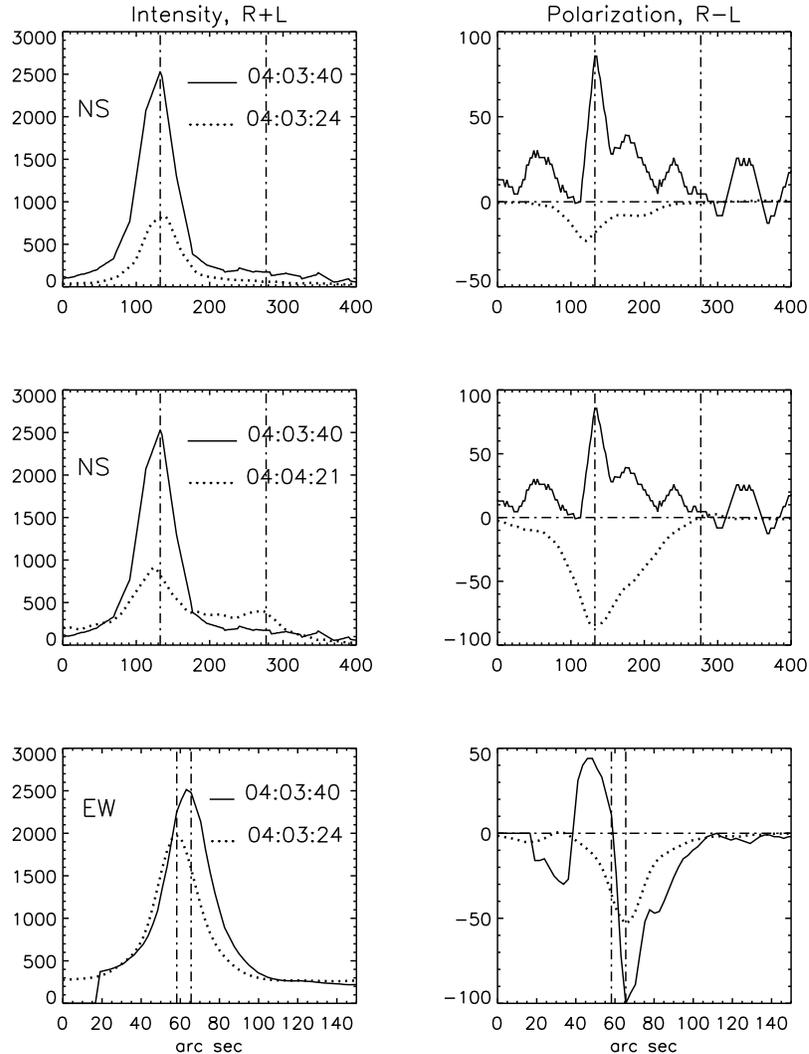} \caption{One-dimensional brightness
distributions (scans) recorded with the EW and NS arrays at 5.7~GHz
in intensity (left) and polarization (right). Each point in these
brightness distributions is a result of integration of the true
brightness along a line parallel to the dash-dotted EW (EW-scans) or
NS (NS-scans) lines shown in Figure~\ref{Fig2_3}. Solid profiles
correspond to the burst peak. Profile magnitudes are in arbitrary
units. Vertical dash-dotted lines correspond to positions marked in
the NoRH map in Figure~\ref{Fig2_3} (left). Middle panels show the
appearance of the unpolarized RS in the NS scans. \label{1dssrt}}
\end{figure}

\begin{figure}
\epsscale{0.9} \plotone{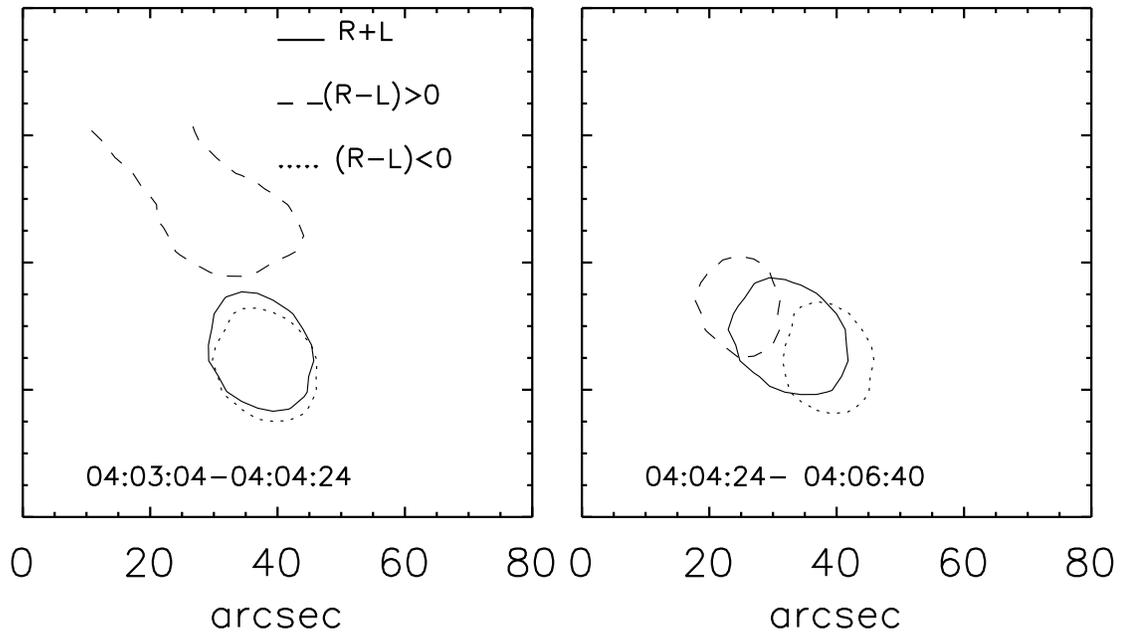} \caption{Brightness distributions in
intensity and polarization at 17~GHz. Average values of contours are
$4\times 10^7$~K in intensity, $1.5\times 10^4$~K in right handed
polarization, and $-6\times 10^6$~K in left handed polarization in
the left panel, and $ 2\times 10^6$~K in intensity, $ 1.35\times
10^4$~K in right handed polarization, and $-10^5$~K in the right
panel. \label{pol_mag}}
\end{figure}

The only transport model allowing radio to lead  HXR is a 'single
reflection' model, which is adopted below. Specifically, if a
particle beam with some angular scatter is injected at a asymmetric
magnetic trap towards a foot-point with stronger magnetic field,
most of the electrons can be reflected back to form a hollow beam,
produce gyrosynchrotron emission in the region of relatively strong
magnetic field, and then after corresponding travel time over the
loop reach the other foot point with weaker magnetic field to
penetrate deeper into the chromosphere and produce HXRs. In this
case the observed delay between radio and HXR originates naturally.
In addition, the presence of initially downward injected beam is
confirmed by the reverse drifting coherent subbursts (Figure
\ref{fine}) leading  both microwave and HXR peaks by a fraction of
second.

In our event the formation of a asymmetric loop is likely because
the photospheric magnetic field has the extremes of $-170$ G and
+340 G at the flare kernels K1 and K2 respectively. This means that
the magnetic field in the radio source should belong to the range
+170~G~$ < B <$ +340~G. Indeed, if the regions of weaker magnetic
field provided noticeable radio emission, then the trapping of the
particles between $-170$ G and 170 G loop layers were important,
which is not observed.

Third, the information about the largest possible value of the
magnetic field at the source allows to make a firm conclusion about
possible role of the gyrosynchrotron self-absorption in the event.
This question is important because  the low-frequency slope of the
spectrum ($\gamma= 1.9$) is consistent with that expected for the
optically thick gyrosynchtotron radiation \citep{Dulk_1985}.
However, with the given magnetic field range and the given electron
distribution it is impossible to ensure the spectral peak at about
17 GHz (as observed) by the self-absorption effect unless the source
is extremely compact, with the linear scale less than $700$ km. In
this case, however, the number density of the fast electrons will
exceed $10^{12}$ cm$^{-3}$, which we believe is not realistic. Thus,
we adopted the typical angular scale of the source to be 6''
consistent with imaging observations of the radio source.

Alternatively, a large value of the microwave spectral peak
frequency can be provided by Razin effect, which requires high
plasma density at the radio source. Indeed, the presence of high
plasma density at the source is likely because we observe type III
like drifting bursts around 5 GHz (Figure~\ref{fine}).
Accordingly, we adopt the background plasma density to be
\begin{equation}\label{density_fixed}
 n_e = 3\cdot 10^{11} {\rm cm}^{-3}.
\end{equation}
This estimate agrees with the value of the emission measure
determined from the soft X-ray emission \citep{Uddin04}.

In such a dense plasma the Razin-effect is strong for the entire
range of the magnetic field $170-340$ G. In conditions of strong
Razin effect the low-frequency slope of the gyrosynchrotron spectrum
in the uniform source  is much steeper than one observed. We must,
therefore,  conclude that the low-frequency slope of the spectrum is
eventually formed by the source inhomogeneity, i.e., by different
layers with the magnetic field ranging from 170 G to 340~G in the
region of kernel K2. In such a dense plasma the free-free absorption
is typically important throughout the microwave range
\citep{Bastian_etal_2007}. However, high plasma temperature of about
3$\cdot10^7$K was determined for this event from the soft X-ray data
\citep{Uddin04}, therefore, the free-free optical depth is less than
unity at $f>7$ GHz. Thus, we will not take into account the
free-free absorption in our analysis.

We note that the spectral peak provided by the Razin effect
increases as the magnetic field decreases. This means, in
particular, that \emph{lower-frequency} emission should arise
\emph{lower} in the loop, in contrast to usual situation when
\emph{higher-frequency} sources are located lower in the corona. We
checked that the position of the brightness peak at 17~GHz is indeed
displaced southwest by about 5'' relative to the 34~GHz brightness
peak in agreement with the prediction made. Therefore, the
high-frequency emission should arise from the region of the lowest
possible magnetic field, thus, we adopt
\begin{equation}\label{B_fixed}
    B = 180~G
\end{equation}
consistent with the requirement $B > 170$ G. Now, when  most of the
source parameters are fixed based on straightforward use of various
observational indicators, we can turn to formulating the forward
fitting model.

\subsection{Forward fitting scheme}

Even though there are many individual measurements of the radio
emission produced by the considered event, we can make use of only a
minor fraction of them. Indeed, since the low-frequency part of the
spectrum is related to the inhomogeneity of the source, which cannot
be reliably constrained by the observations, we can only model the
high-frequency part of the radio spectrum by the uniform source
(which we refer to as 'high-frequency source').

Specifically, we adopt that the high-frequency source is entirely
responsible for the emission at 80~GHz and 35~GHz and for a
significant fraction of the emission at 17~GHz. Therefore, we have
at best five different observational data points (Stokes I and V at
17~GHz and 35~GHz and Stokes I at 80~GHz), which may allow for
finding four free parameters or less. It is clear, that the weights
of the measurements are different from each other: the highest
weight is given to measurements at 35~GHz (5$\%$ error in the
intensity, and 10$\%$ error in polarization) as they have a small
experimental error and should be well described by the uniform
source model. The 80~GHz intensity has lower weight (while it should
be described by the uniform model even better that the 35~GHz data,
the 80~GHz intensity is measured with 40$\%$ error). The
experimental error at 17~GHz is small, however, the effect of source
inhomogeneity becomes important. Thus, we adopt that the
high-frequency source provides  70$\%$ of the observed flux at
17~GHz with  error of 40$\%$; the error of 100$\%$ in the degree of
polarization at 17~GHz is adopted.

In previous subsection we mentioned already that one of the free
parameters we determine from the fitting  is the characteristic life
time $\tau_l$ of the emitting electrons at the radio source. Another
important parameter, which is not known from the observations, is
the viewing angle $\theta$ between the line of sight and the
direction of the magnetic field at the source, so the viewing angle
is the second free parameter in the forward fitting scheme. Thus, we
have to use a test function for the angular part of the electron
distribution, which depends on only one or two free parameters. Our
model involving one reflection of the electrons from the magnetic
mirror cannot be described by a function with a single free
parameter: it must include both the direction where the angular
distribution reaches the maximum, which differs from the direction
along the field lines after the reflection, and typical angular
scatter of the distribution. Thus, a test function with two free
parameters is necessary. As a simplest approximation we adopt a
normalized gaussian angular distribution over the cosine of the
pitch-angle with unknown mean and dispersion:
\begin{equation}
\label{beam_mu_gauss}
    f_2(\mu) \propto
\exp\left(- {(\mu-\mu_0)^2 \over \Delta \mu^2}\right) .
\end{equation}

Then, we apply a nonlinear code that adjusts model free parameters
to minimize the $\chi^2$ statistics using the downhill simplex
method \citep{Press_etal_1986}. The $\chi^2$ statistics is
calculated as
\begin{equation}
\chi^2 = %{1\over {N-n}}
\sum_{i=1}^N {{(S_i^{obs}-S_i^{mod})^2}\over{\sigma_i^2}},
\end{equation}
where $S_i^{obs}$ are the observational data of either Stokes I or V
for the selected three frequencies, $\sigma_i$ are defined by the
errors introduced above, $S_i^{mod}$ are the model values of the
intensity and polarization.

Our current forward fitting scheme is different from the scheme
applied previously   \citep{Bastian_etal_2007} in two instances.
First of all, minimizing the $\chi^2$ statistics we employ both
intensity and polarization data simultaneously in the same run,
which is the first example when the polarization data is used for
quantitative diagnostics of the fast electron distribution. And
then, our model function is the exact expression of the
gyrosynchrotron emission including the summation over the series of
the Bessel functions and their derivatives. Because the magnetic
field at the source is somewhat low, the contribution of large
harmonics is important at high frequencies, therefore, we had to use
the sum over 1600 terms of the series to describe the emission
correctly up to 80~GHz. This made our scheme computationally
expensive: the full run with one spectrum took about 50 hours of the
PC with 2.1 GHz processor.

\subsection{Forward fitting results}

\begin{figure}
\epsscale{1.0} \plotone{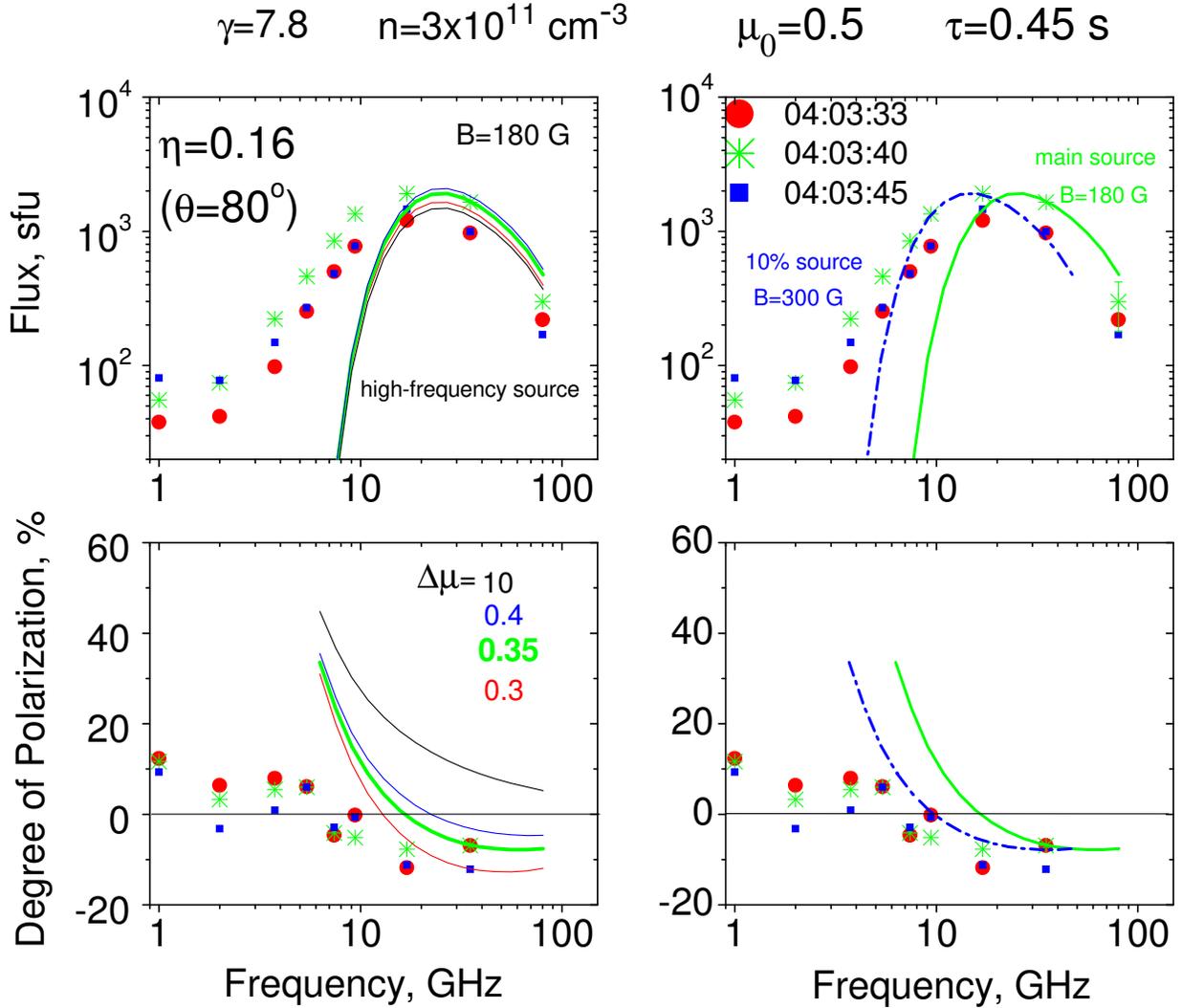} %{Gyro_fit_full.eps}
\caption{Observed spectra at various time frames (symbols) and
fitting results (curves). Left column presents the best fit curves
for the total intensity and degree of polarization for the
high-frequency source, green/thick lines as well as curves
obtained by small variations of the dispersion in the fast
electron angular distribution (blue and red curves). Black curve
show the results for the isotropic electron distribution all other
conditions being equal. Right column displays the best fit curves
for the high-frequency source (green/solid) supplemented by curves
for the emission from a secondary source with higher magnetic
field and smaller amount of the fast electrons (blue/dash-dotted
curves) all other conditions being equal. This simple two-source
model is good enough to describe the total intensity, although
insufficient to fit the polarization data. \label{gyro}}
\end{figure}

Given that the observed spectrum does not evolve much during the
burst, while the forward fitting scheme employing exact
gyrosynchrotron equations is very time consuming, we concentrated on
study of the emission at the peak time of the burst (04:03:40~UT)
only. Result of the fitting is shown in Figure~\ref{gyro}. A number
of important things should be noticed in the figure. First, the
model radio spectra (top left panel) obtained for the electron
energy spectrum derived from HXR data are good match for the
observed high-frequency part of the radio spectra. Thus, the data is
consistent with the model assumption of a single power-law electron
spectrum in the event. Second, the curves for the degree of
polarization are very sensitive to the details of the electron
angular distribution. In particular, the polarization data is
entirely inconsistent with isotropic angular distribution of the
fast electrons, since the isotropic distribution would produce
X-mode polarized emission, while the observed polarization
corresponds to O-mode emission. The gyrosynchrotron radiation
produced by oblique beam observed by quasitransverse direction is
O-mode polarized as needed. The exact value of the degree of
polarization depends strongly on the details of the pitch-angle
distribution, thus, the joint use of the intensity and polarization
measurements is indeed a key to constrain the angular distribution
of the electron beam. For the peak time of the burst the following
parameters provide the best fit to the observed spectrum and
polarization:
\begin{equation}\label{fit_res}
    \tau_l=0.45~s, \qquad \theta=80^o, \qquad  \mu_0=0.5  , \qquad
    \Delta\mu=0.35.
\end{equation}

All these numbers look reasonable against observations and theory of
gyrosynchrotron radiation from anisotropic electron distribution.
Indeed, the life-time of the fast electrons is small enough, $\tau_l
\approx 0.45$~s, i.e., less than the radio peak duration as
required. Then, as is known from the gyrosynchrotron theory, the
O-mode polarization of the optically thin source is only possible
for beam-like electron distributions and for viewing angles larger
than the peak angle in the pitch-angle distribution. The obtained
values of $\theta$ and $\mu_0$ obey these requirements since
$\cos\theta=0.16 < \mu_0$.

It is tempting now to subtract the model contribution of the
high-frequency source from the observational data points and repeat
the forward fitting procedure. Indeed, we can perform a reasonable
fit to the total power data. As an example, a contribution of a
"lower-frequency source" with the same pitch-angle distribution but
with stronger magnetic field ($B \approx 300$ G), while smaller
life-time of the fast electrons ($\tau_l \approx 0.05$~s) is shown
in the figure.

However, it is not possible to perform a consistent fit to the
polarization measurements, which is the key to constrain the
electron angular distribution, because the observed degree of
polarization is essentially the result of averaging of various
contributions along the non-uniform source.  This conclusion is in
agreement with high spatial and temporal variability of the
polarization patterns observed at 5.7~GHz, which is most probably a
result of changing relative contributions from different parts of an
inhomogeneous source (note a very strong frequency dependence of the
degree of polarization in the model curves below 10 GHz in
Figure~\ref{gyro}). Therefore, the low-frequency observations cannot
be conclusively fitted by the uniform source model.

\section{Discussion}

In this paper we presented a new tool of studying electron beams
accelerated during solar flares by analysis of the gyrosynchrotron
emission produced by the beams. Methodologically, this result is
achieved by quantitative use of the polarization measurements of the
microwave gyrosynchrotron emission. Specifically, to obtain the
information of the beam-like angular  distribution of the
accelerated electrons we developed a nonlinear $\chi^2$ fit
employing Stokes I $\&$ V measurements simultaneously and exact
gyrosynchrotron formulae.

This approach allows for unambiguous detection of oblique beams at
the radio source. In addition, our scheme yields a number of
important physical parameters of the radio burst such as the viewing
angle of the radio emission relative to the magnetic field at the
source, characteristic parameters of the electron distribution over
pitch-angle, and typical life time of the electrons in the radio
source, see Eq. (\ref{fit_res}).

The diagnostics of the electron beam obtained from the fit of the
gyrosynchrotron data within 'one-reflection model' is consistent
with all other available observations. In particular, the time
delays of the hard X-rays relative to the microwaves (a fraction of
second) is consistent with the transit time of the electrons with
relatively large pitch-angles (found from the forward fitting
technique) through the flaring loop of about 10$^4$ km in
projection. In addition, the fine structures consisted of the
reverse drift bursts followed by the normal drift bursts detected at
about 5~GHz are in agreement with the 'one-reflection model', which
implies the downward beam propagation followed by its reflection at
the magnetic mirror and consequent upward motion of the beam. The
radio data suggest that the main radio source is filled by a rather
dense plasma, Eq. (\ref{density_fixed}), with the plasma frequency
of about 5 GHz. The radio emission observed at lower frequencies
(3.75 and 2 GHz) cannot be produced at this main source. The most
straightforward interpretation for this low-frequency component is
to postulate an adjacent more tenuous loop, where a minor fraction
of the accelerated electrons produce lower-frequency radiation. Such
a larger loop is likely and confirmed by the presence of a remote
source magnetically connected with the main flare site. If so,
larger decay constant and the observed delay of the low-frequency
emission relative to higher-frequency emission (see \S 3.1) receives
a natural interpretation because it simply relates to larger source
size implying longer transit time, where, in addition, some trapping
of the electrons can be important.

Another possible effect of the dense plasma is an enhanced role of
the Coulomb collisions on the radiating electrons. Given relatively
low value of the magnetic field at the source, it is easy to
estimate the typical energies of the radio emitting electrons as
$1-10$ MeV, which have the Coulomb energy decay time $\tau_c > 15$
s, which is much longer than the transit time. The time of
isotropization of these relativistic electrons due to Coulomb
collisions is even longer than the energy decay time
\citep{Petrosian_1985}. We, thus, conclude that the Coulomb
collisions have very little effect on the electron distribution at
the time scale of the electron precipitation ($\sim 1$ s) in this
event.

Although we quantitatively  describe the emission at the burst peak
only, we can reliably extrapolate the main findings, such as
acceleration of the fast electrons in the form of beams and the
one-reflection transport model to the entire burst duration. It
follows from the close similarity between the radio and the hard
X-ray light curves and from the constancy of the sense of
polarization in the main source at 17 GHz. This means that some of
the flares, such as one considered here, can predominantly
accelerate electrons along the magnetic field lines even though the
pitch-angle distribution of the beams has some angular scatter.

The potentiality of the developed method is very strong and will be
especially helpful when imaging spectroscopy data is availably.
However, the routine use of this method requires optimization of the
computation scheme, which is not fast enough at present. Therefore,
deducing simplified, although precise enough, gyrosynchrotron
formulae for anisotropic electron distributions can be very helpful
here.

\section{Conclusion}

Electron beams with some angular scatter can efficiently produce
microwave continuum bursts via gyrosynchrotron mechanism. As was
shown theoretically by \cite{Fleish03}, the gyrosynchrotron emission
produced by electron beams can be distinguished from that produced
by isotropic electron populations by analysis of the degree of
polarization of the microwave emission. Here we presented a
compelling example of such event, where the optically thin
gyrosynchrotron radiation is indeed O-mode polarized as expected for
the beam-like distributions. Remarkably, the presence of the beam
is confirmed by the whole set of the available data for this event.

In case of the radio data with good enough quality (including both
intensity and polarization) the use of forward fitting inversion
methods allows for quantitative diagnostics of the fast electron
angular distribution as well as a number of other important physical
parameters of the flaring source. These methods will become much
more useful when the imaging spectroscopy data is available.

\acknowledgments The authors express appreciation to Prof. Kiyoto
Shibasaki and Prof. Hiroshi Nakajima  for the NoRP calibration
corrections they provided. This work was supported in part by NSF
grants ATM-0607544 and ATM-0707319  to New Jersey Institute of
Technology,  by the Russian Foundation for Basic Research, grants
No. 06-02-16295, 06-02-16859, 06-02-39029, 07-02-01066, NFSC project
No. 10333030 and ''973'' program with No. 2006CB806302. We have made
use of NASA's Astrophysics Data System Abstract Service.

\end{document}